\documentclass[seceq]{ptptex}
\usepackage{wrapft}
\usepackage{graphicx}

\title{ Nuclear Matrix Element for Two Neutrino Double Beta Decay From $^{136}$Xe}

\author{
Hiroyasu \textsc{Ejiri}$^{1,}$\footnote{ejiri@rcnp.osaka-u.ac.jp} 
}

\inst{
$^1$  Research Center for Nuclear Physics, Osaka University, Osaka 567-0047, Japan,\\
Nuclear Science, Czech Technical University, Brehova, Prague, Czech Republic,
}

\abst{
The nuclear matrix element for the two neutrino double beta decay (2$\nu \beta \beta $) of $^{136}$Xe $\rightarrow ^{136}$Ba was evaluated by FSQP(Fermi Surface Quasi Particle model), where  experimental GT strengths measured by the charge exchange reaction and those by the $\beta $ decay rates were used. The 2$\nu \beta \beta $ matrix element is given by the sum of products of the single $\beta ^- $ and $\beta ^+$ matrix elements via low-lying (Fermi Surface) quasi-particle states in the intermediate nucleus. $^{136}$Xe is the semi-magic nucleus with the closed neutron-shell, and the $\beta ^+$(proton$\rightarrow $neutron) transitions are almost blocked. Thus the 2$\nu \beta \beta $ is much suppressed. The evaluated 2$\nu \beta \beta $ matrix element is consistent with the observed value.}

\begin{document}
\maketitle

Neutrino-less double beta decays(0$\nu \beta \beta $), which violate the lepton number conservation law by $\Delta L=2$, are beyond the standard electro-weak model(SM). Studies of 0$\nu \beta \beta $ are of great interest to investigate the Majorana property of neutrinos, the absolute neutrino mass scale and other fundamental properties of neutrinos and weak interactions beyond SM, as discussed in review articles and references therein \cite{eji05,avi08,eji10,ver12}. Here 0$\nu \beta \beta $ nuclear matrix elements($M^{0\nu}$) are crucial to extract the neutrino mass and other properties of particle physics interests from 0$\nu \beta \beta $ experiments, as discussed in the review articles and others \cite{vog86,suh98,eji00,rod07,sim01,civ05,hor10,bar09}.
 
Two neutrino double beta decays(2$\nu \beta \beta $) are within SM, and their matrix elements ($M^{2\nu }$) are derived from 2$\nu \beta \beta $ experiments. They provide useful information on nuclear structures for evaluating $M^{0\nu}$. Nuclear models and nuclear parameters for $M^{0\nu}$ are necessarily required to be consistent with those for $M^{2\nu }$.

Recently single $\beta ^-$ strengths for $^{136}$Xe were measured by the charge exchange reaction at RCNP Osaka \cite{pup11}, and the $M^{2\nu }$ for $^{136}$Xe was measured by the EXO collaboration \cite{ack11}. 

The present letter aims to show that the FSQP model\cite{eji96,eji09} based on the observed single $\beta $ matrix elements \cite{pup11,son03,son04} for low-lying states reproduce the observed $M^{2\nu }$ for $^{136}$Xe \cite{ack11}, which is the key nucleus for double beta decay experiments. In fact two groups are studying 0$\nu \beta \beta $ rates by using large amount of enriched $^{136}$Xe isotopes \cite{gra08,efr11}. 

The $M^{2\nu }$ values have been measured in many nuclei. They are quite small, almost 2 orders of magnitude smaller than single quasi-particle values. Recently, the 2$\nu \beta \beta $ rate for $^{136}$Xe was measured to be $M^{2\nu }$=0.01 $m_e^{-1}$ \cite{ack11}, which is even smaller by one order of magnitude than those for other nuclei such as $^{100}$Mo, $^{82}$Se, and $^{76}$Ge \cite{ eji05,bal06}. $^{136}$Xe is the semi-magic nucleus with the closed neutron-shell, and the $\beta ^+$(proton$\rightarrow $neutron) transitions are almost blocked, as shown in Fig.1. Thus the 2$\nu \beta \beta $ is much suppressed.

\begin{figure}[h]
\begin{center}
\includegraphics[width=0.8\textwidth]{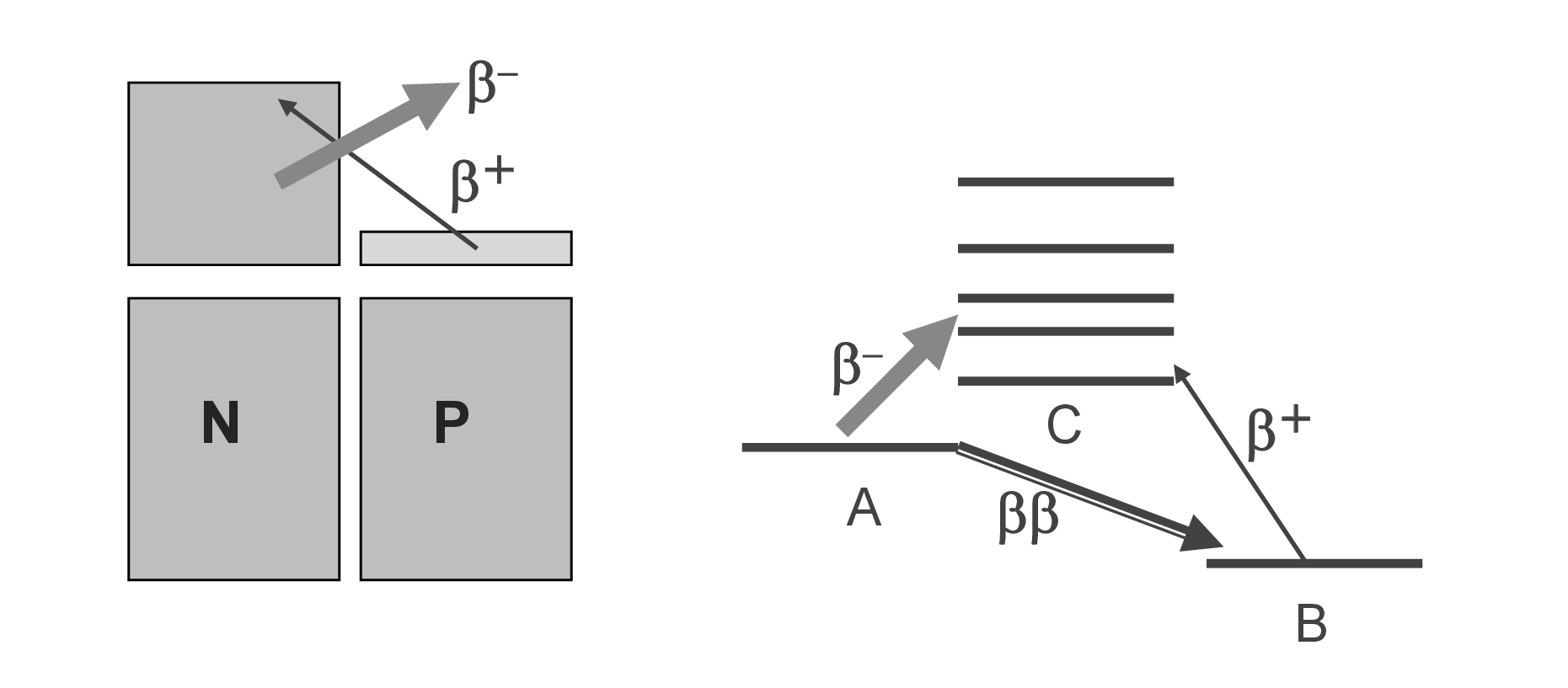}
\caption{(Color online). Energy and transition schemes for the 2$\nu \beta \beta $ decay in the closed neutron-shell nucleus of $^{136}$Xe. $\beta ^-$ transitions from the closed    neutron-shells to vacant proton-shells are favored, but $\beta ^+$ ones from the almost vacant proton-shells to the almost occupied neutron-shells are much suppressed.
}
\label{fig:1}
\end{center}       
\end{figure}

QRPA models \cite{vog86,suh98,rod07,sim01,civ05} show that the small $M^{2\nu }$ values have been explained by the cancellation of the two terms at a certain value of the $g_{pp}$ parameter, which is adjusted so as to reproduce the data \cite{rod07}. It is a challenge for FSQP to reproduce the extremely small $M^{2\nu }$ for $^{136}$Xe by using the experimental single $\beta ^{\pm}$ strengths without adjusting artificially any nuclear parameters.  

$M^{2\nu }$ for 0$^+(N,Z)\rightarrow 0^+(N-2,Z+2)$ is expressed as the sum of the products of the successive single $\beta ^\pm $ matrix elements via low lying quasi-particle states in the intermediate nucleus \cite{eji09}, 
\begin{equation}
M^{2\nu }~=~\sum _{ij} M^-_{ij} M^+_{ij}~\Delta _{ij}^{-1}
\end{equation}
\begin{equation}
 M^-_{ij} ~=~g^{eff}_-~V_n^i(N)U_p^j(Z)~m_{ij},
\end{equation}
\begin{equation}
M^+_{ij}~=~g^{eff}_+~U_n^i(N-2)V_p^j(Z+2)~m_{ij}, 
\end{equation}
where $m_{ij}$ is the reduced GT matrix element for the single particle $i \rightarrow j$ transition, $V_n^i(N)$ and $U_p^j(Z)$ are the occupation and vacancy amplitudes for the 
$i$ th neutron and the $j$ th proton, and so on, and $\Delta _{ij}$ is the energy denominator. Effects of nuclear spin isospin and ground state correlations and nuclear medium effects on the nuclear matrix elements are given by the effective coupling constant $g^{eff}$ in unit of $g_A$ = 1.256. The $V$ and $U$ amplitudes are single quasi-particle values obtained from the BCS equations, while $g^{eff}_{\pm}$ is obtained from the relevant experimental values. 

It is noted here that $M^-_{ij}$ and $M^+_{ij}$ have the same matrix element of $m_{ij}$, and thus their signs are the same. Consequently, their products are all positive, and the $M^{2\nu }$ is a constructive sum of the products. 

The high precision experiment of the charge exchange reaction on $^{136}$Xe was made to study the GT$^-$ strength distribution in the intermediate nucleus of $^{136}$Cs \cite{pup11}.
The running sum of the obtained strengths is shown to compare with the FSQP value as a function of the excitation energy $E_{ij}$ in Fig. 2. The summed strength for the low-lying states is as large as 0.71, reflecting the large neutron occupation amplitudes($V_n^i \approx 1$) for the closed neutron-shell $^{136}$Xe. 
The effective coupling constant is derived as $g^{eff}_-$=0.13, in unit of $g_A$ = 1.256, from the observed 
matrix element of $M^-_{11}$ = 0.39 for the lowest intermediate state(2d$_n$(3/2)2d$_p$(5/2)) at $E_{11}$=0.59 MeV.

The matrix elements $M^-_{ij}$ are evaluated by using the eq.(2) and $g^{eff}_-$=0.13. The running sum of the GT strengths, $\sum B(GT^-)_{ij}=\sum|M^-_{ij}|^2$, is shown in Fig.3. The obtained sum of 0.77 is close to the observed one of 0.71 \cite{pup11}.

The effective coupling constant for the $\beta ^+ $ transition is derived as $g^{eff}_+$=0.15, in unit of $g_A$ = 1.256, from the neighboring $\beta ^+$ decays of $^{138}$Nd and $^{138}$Ce \cite{son03,son04}. By using this coupling constant, the matrix elements of $M^+_{ij}$ were evaluated by means of eq.(3). The running sum of $\sum B(GT^+)_{ij}=\sum|M^+_{ij}|^2$ is obtained as shown in Fig. 3. The strength is mainly at the two lowest states. The strengths at higher states are very small because of the small vacancy probabilities of deep neutron-hole states and the small occupation probabilities of high-lying proton-particle states.

The 2$\nu \beta \beta $ matrix elements are evaluated by using eq.(1) with $g^{eff}_- $=0.13 and $g^{eff}_+$=0.15. The running sum of the matrix elements is shown in Fig. 4 (a:thick solid line).  Fig.4 b:thick dot-dash line is the value with the effective coupling constant $g^{eff}_- $ = 0.125 which is derived so as to reproduce the observed $\Sigma B(GT^-)$, while c:thine solid line is the value with the observed $M_{ij}^-$ for the 2 lowest states and the evaluated values with $g^{eff}_-$ = 0.13 for other states. The evaluated values for the running sum of $\Sigma M_{ij}$ are around 0.012 $m_e^{-1}$ in accord with the observed value of $M^{2\nu }=0.010 ~m_e^{-1}$. The extremely small matrix element is reproduced by the present FSQP model
\begin{figure}[h]
\begin{center}
\includegraphics[width=0.8\textwidth]{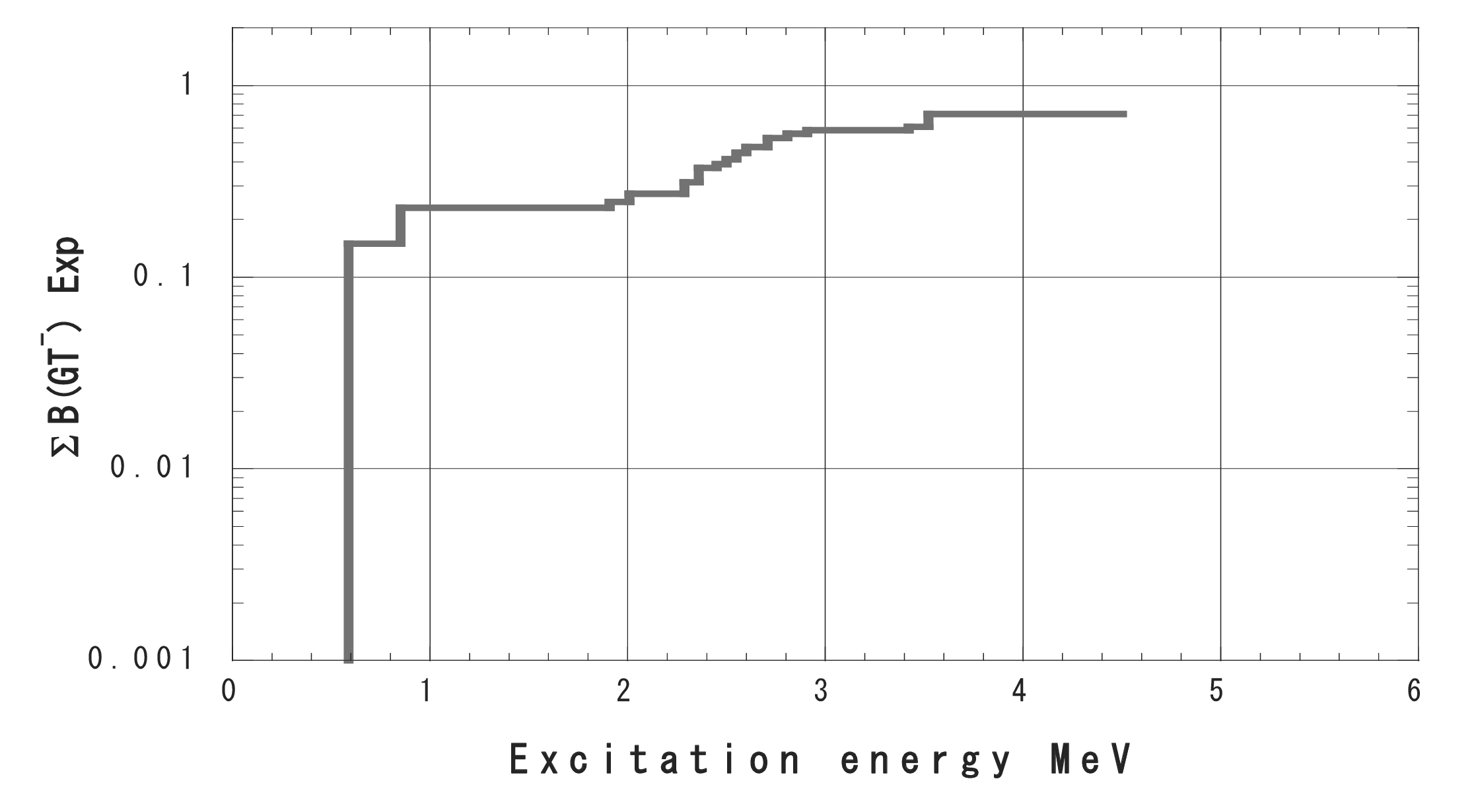}
\caption{(Color online). Running sum of the observed GT$^-$ strengths of $B(GT^-)$ as a function of the excitation energy, as measured by CER \cite{pup11}.}.

\label{fig:2}
\end{center}       
\end{figure}

\begin{figure}[h]
\begin{center}
\includegraphics[width=0.8\textwidth]{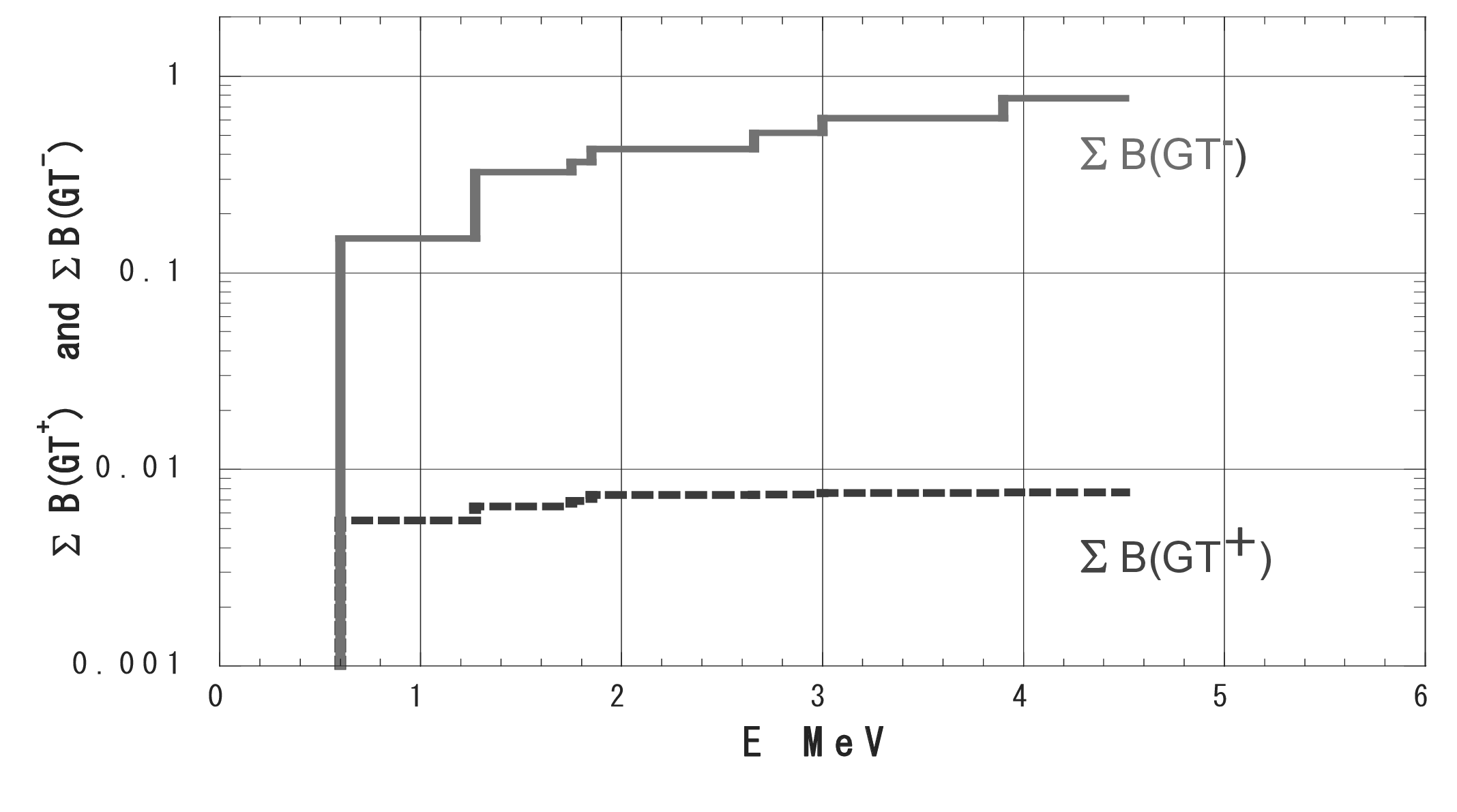}
\caption{(Color online). Running sum of the evaluated GT$^\pm$ strengths of $B(GT^\pm)$ as a function of the excitation energy.}.

\label{fig:3}
\end{center}       
\end{figure}

\begin{figure}[h]
\begin{center}
\includegraphics[width=0.8\textwidth]{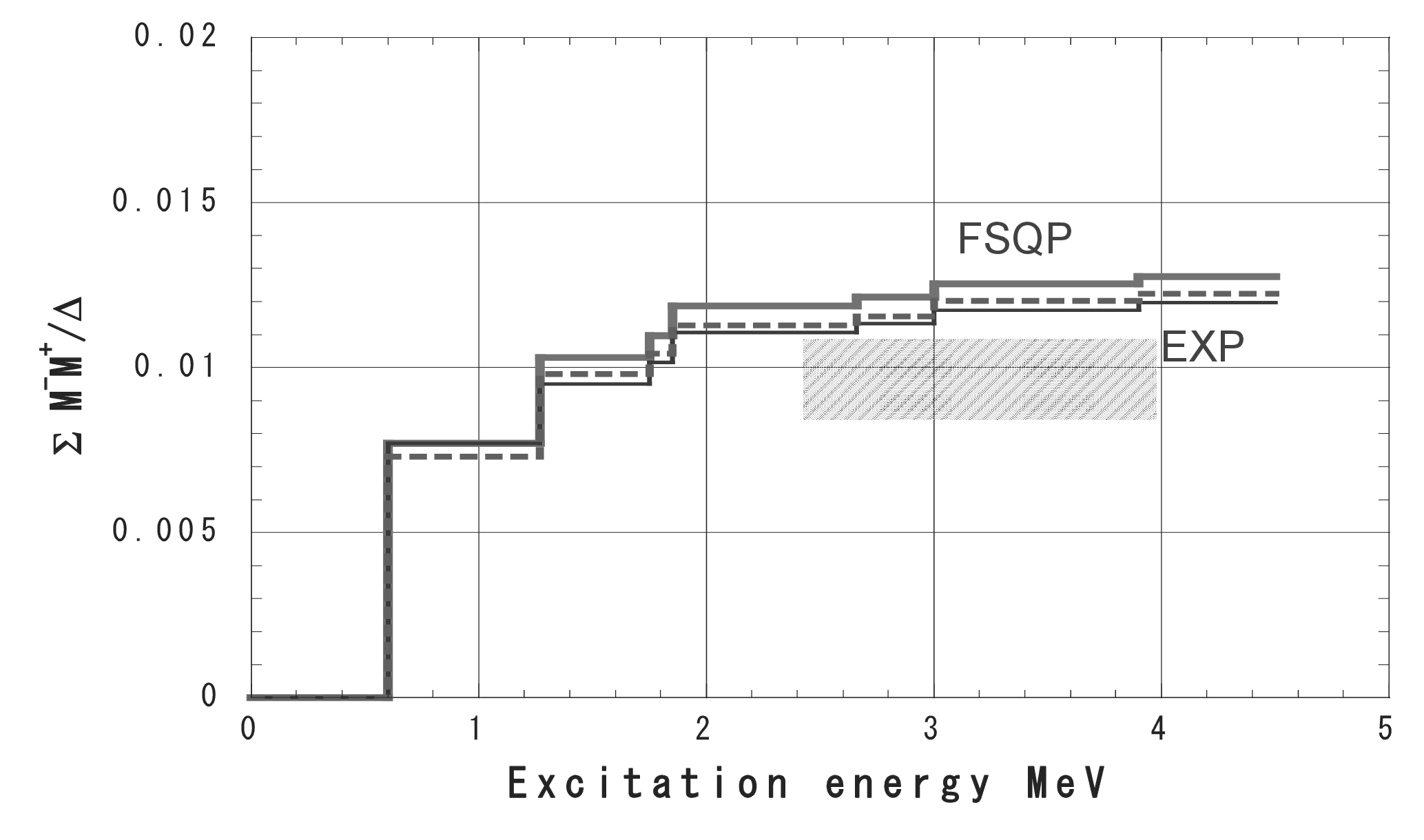}
\caption{(Color online). Running sums of the 2$\nu \beta \beta $ amplitudes of $M^-_{ij} M^+_{ij}~\Delta ^{-1}$ as a function of the excitation energy. The experimental value with 2$\sigma $ is shown by the hutched region. a:thick solid line and b:thick dashed line are the FSQP evaluations with $g^{eff}_-$ = 0.130 and 0.125, respectively. c:thine solid line is the one with the experimental $M_{ij}^-$ values for the 2 lowest $1^+$ states (see text).}.

\label{fig:4}
\end{center}       
\end{figure}

The small $2\nu \beta \beta $ matrix element can be understood as follows. The observed single $\beta ^{\pm}$ matrix elements of $|M^-_{ij}|$ and $|M^+_{ij}|$ in the present mass region of $A$=150 are smaller by the factors $g^{eff}_-$=0.13 and $g^{eff}_+$=0.15, both in unit of $g_A$ = 1.256, than the 
single quasi-particle values. Thus the $M^{2\nu }$ gets smaller by the factor
$g^{eff}_{\beta \beta}$ =$g^{eff}_- \times g^{eff}_+ $0.02 than the single quasi-particle matrix element. This factor is nearly the same as the values in other mass region. It is an universal coefficient, standing for the ground state and spin isospin correlations and nuclear medium effects on GT weak strengths. 

The $\beta ^-$ and $\beta^+$ matrix elements of $M^-_{ij}$ and $M^+_{ij}$ are proportional to the $U$ and $V$ factors of $V_n^i(N)U_p^j(Z)$ and $U_n^i(N-2)V_p^j(Z+2)$, respectively. In case of $^{136}$Xe with $N$=82 and $Z$=54, the neutron shell is full (closed) and the proton shell is almost vacant. 
Then, the factor for $M^-_{ij}$ is $V_n^i(N)U_p^j(Z) \approx 1$, and 
 the factor for $M^+_{ij}$ is $U_n^i(N-2)V_p^j(Z+2) \approx 0.05 \ll1$. The $\beta ^+$ transitions from nearly vacant proton-shells to nearly occupied neutron-shells are much suppressed.    
Consequently, the product of the $U$ and $V$ factors for the $\beta ^-$ and $\beta ^+$ matrix elements becomes as small as 0.05, resulting in the small $M^{2\nu}$ matrix element. 

The  $U$ and $V$ factors($UV$) for transitions between half-filled shells are around 0.5 for both $\beta ^-$ and $\beta ^+$ transitions and the product is around 0.25. Then $M^{2\nu}$  for the semi-magic $^{136}$Xe is smaller by a factor around 5 than the factors for other nuclei such as $^{76}$Ge and $^{82}$Se.

The evaluated 2$\nu \beta \beta $ is a bit larger than the observed value
 by 20 $\%, 
\Delta M^{2\nu} \approx 0.002$. 
This is partly due to the uncertainty of the $g^{eff}_{\pm}$. 
In particular, the matrix elements of $M_{ij}^+$ for excited states are small because of the small $U$ and $V$ factors. Then the ground state correlation contributes destructively to the matrix element, and thus $g^{eff}_+$ for the excited small gets smaller. Actually, the running sum of the evaluated matrix elements up to 1.8 MeV is around 0.10 in good agreement with the observed value. The GT giant resonance may get relatively important in such cases with very small matrix elements for the low-lying states. Nevertheless the contribution itself is as small as $\Delta M^{2\nu} \approx  0.002$. 

\begin{figure}[h]
\begin{center}
\includegraphics[width=0.5\textwidth]{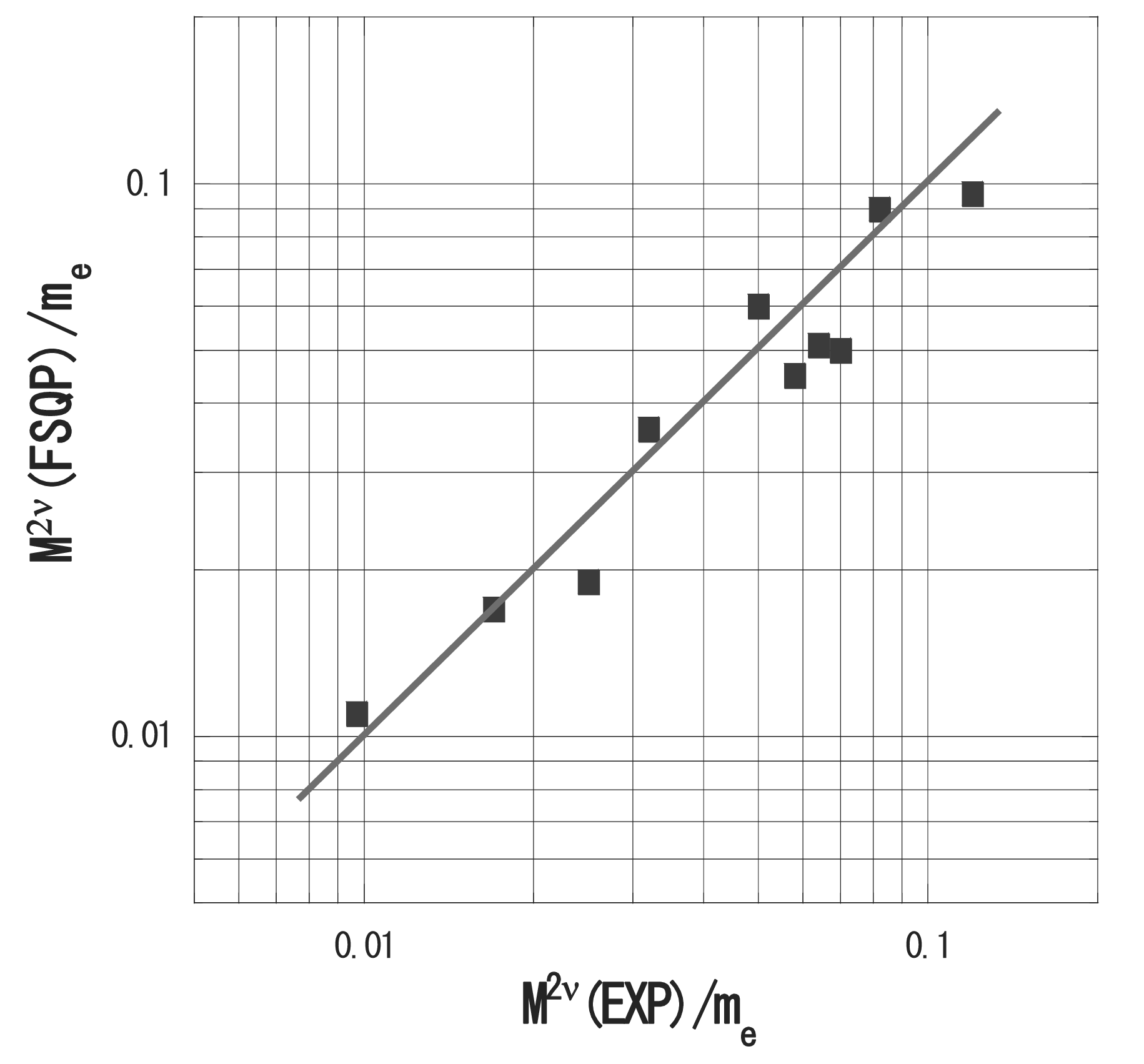}
\caption{(Color online). Experimental 2$\nu \beta \beta $ matrix elements (x axis) and the evaluated ones (y axis) in terms of FSQP. The smallest values are for $^{136}$Xe}.

\label{fig:5}
\end{center}       
\end{figure}

\begin{figure}[h]
\begin{center}
\includegraphics[width=0.7\textwidth]{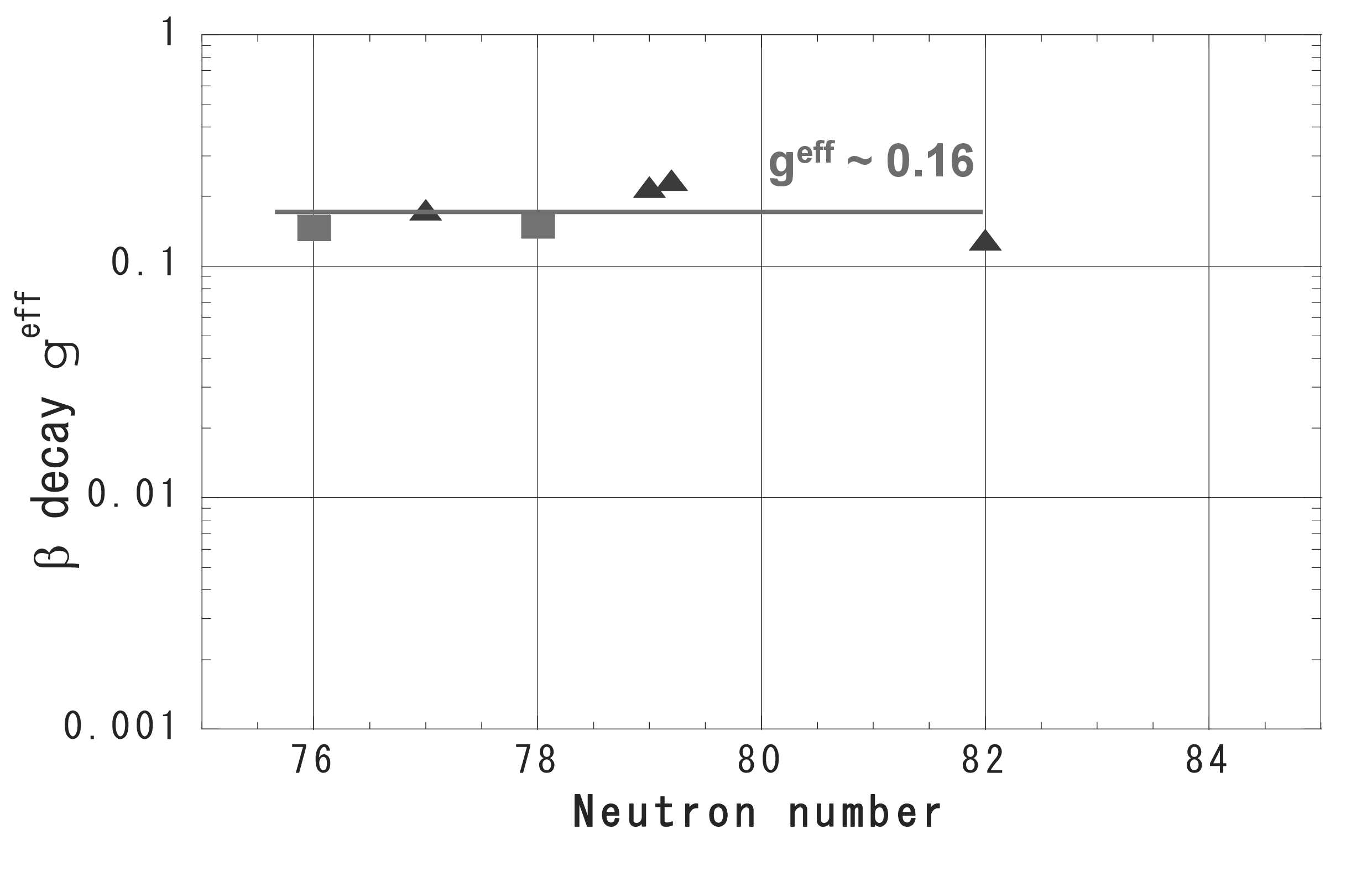}
\caption{(Color online). Macro 6Effective axial vector weak coupling constants $g^{eff}_{\pm}$ in unit of $g_A$ = 1.256 for 0$^+ \rightarrow 1^+$ GT decays for nuclei in the present mass region. Squares and triangles are for $\beta ^-$ and $\beta ^+$ decays, respectively.}.

\label{fig:6}
\end{center}       
\end{figure}

The observed $2\nu \beta \beta $ matrix elements are compared with the FSQP evaluations based on  the experimental $\beta ^- $ and $\beta ^+$ matrix elements.
The FSQP values are in agreement with the observed values, as shown in Fig.5. The small matrix elements of $M^{2\nu}$ are mainly due to the core effects of the spin isospin correlation and nuclear medium, which are represented by
 $g^{eff}_- \times g^{eff}_+ $ $\approx $0.02.
 Most $\beta \beta $ strengths are absorbed into higher excited states such as the $\beta \beta $ giant resonances and isobars. 

The nuclear surface and structure effects are represented by the $U$ and $V$ factors. The factors become small in semi-magic and nearly semi-magic nuclei, where the neutron shells are almost closed and the proton shells are almost vacant. The very small 
$M^{2\nu}$ for $^{136}$Xe can be reproduced by the small $U$ and $V$ factors for $M^+_{ij}$.
The vary small values of $M^{2\nu}$  and the isotope dependence are not necessarily due to the accidental cancellations of the two terms at the $g_{pp} \approx 1$.

Let us compare the single $\beta $ decay rates and the present $g^{eff}_{\pm}$ values with those in neighboring nuclei. The log$ft_-$=4.42 for $^{136}$Xe is similar as log$ft_- \approx$4.2 in the neighboring nuclei, while the log$ft_+$ for the $\beta ^+$ decays in the mass region is around 5.2. The $\beta ^-$ rates are larger by an order of magnitude than the $\beta ^+$ rates. This is  because the $U$ and $V$ factors for $\beta ^-$ decays are much larger than those for the $\beta ^+$ decays. The $g^{eff}_-=0.13$ and $g^{eff}_+ = 0.15$ for $^{136}$Xe are same as those in other nuclei, as shown in Fig. 6. The uniform values of $g^{eff}_\pm \approx 0.16$ for both $\beta ^-$ and $\beta ^+$ decays do indicate that the $g^{eff}$ stands for the universal core effects and the nuclear structure/surface effects are well expressed by the $U$ and $V$ factors.

 It is important to note that the present arguments for GT 1$^+$ matrix elements may be extended to 2$^-$ and higher multi-pole ones for 0$\nu \beta \beta $ matrix elements. They might be much reduced by the nuclear core effects and the nuclear surface ($U$ and $V$ factors) effects as in case of $2\nu \beta \beta $ matrix elements. This would contradict with commonly believed theoretical predictions that the 0$\nu \beta \beta $ matrix elements would be all large and nearly the same in all nuclei, in contrast to 2$\nu \beta \beta $ ones. 

Then experimental studies of single $\beta $ matrix elements and nuclear structures relevant to DBD matrix elements are encouraged \cite{eji00,eji10,aki97,fle08,zeg11}.  

The author thanks Prof. D. Frekers, Prof. M. Harakeh and Prof. F. Iachello for valuable discussions.

\end{document}